# Bridging the Gap Between Security Metrics and Key Risk Indicators: An Empirical Framework for Vulnerability Prioritization

Emad Sherif[1], Iryna Yevseyeva[1], Vitor Basto-Fernandes[2,3], and Allan Cook[1]

[1]Faculty of Technology, Arts and Culture, De Montfort University, Leicester, United Kingdom
[2]Instituto Universitário De Lisboa (ISCTE-IUL), University Institute of Lisbon, ISTAR-IUL, Lisboa, Portugal
[3]Sorbonne Université, CNRS, LIP6, 75005 Paris, France

Corresponding author: Emad Sherif (e-mail: P2648946@my365.dmu.ac.uk).

**Abstract** Organisations overwhelmingly prioritize vulnerability remediation using Common Vulnerability Scoring System (CVSS) severity scores, yet CVSS classifiers achieve an Area Under the Precision-Recall Curve (AUPRC) of 0.011 on real-world exploitation data—near random chance. We propose a composite Key Risk Indicator grounded in expected-loss decomposition, integrating dimensions of threat, impact, and exposure. We evaluated the KRI framework against the Known Exploited Vulnerabilities (KEV) catalog using a comprehensive dataset of 280,694 Common Vulnerabilities and Exposures (CVEs). KRI achieves Receiver Operating Characteristic Area Under the Curve (ROC-AUC) 0.927 and AUPRC 0.223 versus 0.747 and 0.011 for CVSS (+24%, ×20). Ablation analysis shows Exploit Prediction Scoring System (EPSS) alone achieves AUPRC 0.365, higher than full KRI (0.223), confirming that EPSS and KRI serve distinct objectives: EPSS maximizes raw exploit detection, while KRI re-orders by impact and exposure, capturing 92.3% of impact-weighted remediation value at k = 500 versus 82.6% for EPSS, and surfacing 1.75× more Critical-severity exploited CVEs. KRI's net benefit exceeds EPSS whenever the severity premium exceeds 2×. While EPSS serves as a robust baseline for exploit detection, the KRI framework is the superior choice for organizations seeking to align remediation efforts with tangible risk reduction.

*Index Terms* Cyber risk management, key risk indicators, machine learning, security metrics, vulnerability prioritization.

## I. Introduction

Cyber risk management has emerged as a critical organizational priority in an era characterized by the escalating frequency, sophistication, and economic impact of cyberattacks [1], [2]. The digital transformation of business operations, expansion of cloud computing, and proliferation of interconnected Internet of Things (IoT) devices have dramatically expanded organizations' attack surfaces, making systematic vulnerability management essential. Despite significant investments in cybersecurity infrastructure and personnel, organizations continue to struggle with effective prioritization of remediation efforts, often driven by inadequate risk assessment methodologies [3].

Traditional approaches to vulnerability management have historically relied on Security Metrics (SMs) such as the Common Vulnerability Scoring System (CVSS), a widely adopted framework developed to measure the technical severity of discovered vulnerabilities [4]. CVSS produces a numerical score ranging from 0 to 10, enabling organizations to prioritize remediation efforts. However, the CVSS framework suffers from fundamental limitations: it was designed to capture technical severity independent of real-world exploitation context, environmental factors, and threat landscape dynamics. Consequently, numerous empirical studies have demonstrated that CVSS scores exhibit poor correlation with actual exploitation events in production environments [5], [6].

This disconnect between theoretical severity and practical exploitation likelihood creates a significant operational problem. Security teams, constrained by limited resources and remediation capacity, face the challenge of prioritizing among thousands of discovered vulnerabilities. When prioritization relies primarily on CVSS severity, organizations frequently allocate significant effort to remediate technically severe vulnerabilities that are rarely exploited in practice, while deprioritizing vulnerabilities with lower severity scores that may be actively exploited by adversaries. This misalignment wastes security resources and leaves organizations vulnerable to attacks targeting overlooked but readily exploitable weaknesses.

To address these shortcomings, security researchers and practitioners have increasingly advocated for the adoption of Key Risk Indicators (KRIs); comprehensive metrics that incorporate exploitability assessments, systemic weakness patterns, and contextual risk factors [1], [7], [8]. Among these innovations, the Exploit Prediction Scoring System (EPSS) has emerged as a particularly promising metric, leveraging machine learning and threat intelligence data to estimate the probability that a vulnerability will be exploited within 30 days of public disclosure [9]. The theoretical foundation underlying KRIs is compelling: by combining indicators of exploitation likelihood (EPSS), technical severity (CVSS), and systemic weakness prevalence (CWE), organizations can construct a multidimensional risk profile that aligns technical indicators with adversarial behavior patterns and operational risk priorities [6].

Despite the conceptual advantages of KRIs, empirical validation of their superiority over traditional metrics remains limited [10]. Most prior work has focused on criticizing CVSS limitations rather than demonstrating the predictive power of composite risk indicators [11], [12], [13]. This study aims to bridge this research gap through rigorous quantitative evaluation.

This work pursues three interconnected objectives: to empirically validate that KRIs outperform traditional metrics in predicting real-world exploitation events; to develop and document a reproducible methodological framework for integrating publicly available datasets into a cohesive analytical pipeline; and to introduce a decision-objective framework that empirically resolves the EPSS/KRI trade-off, demonstrating when each metric is optimal through decision-centric analyses including Precision@k, severity-stratified recall, and a severity-weighted remediation simulation, and translating these findings into a practitioner decision guide grounded in expected-loss principles.

We contribute to the literature in three substantive ways; First, we provide quantitative evidence that KRI substantially outperforms CVSS metrics and, under impact-weighted evaluation objectives, outperforms EPSS alone; Receiver Operating Characteristic Area Under the Curve (ROC-AUC) and Area Under the Precision-Recall Curve (AUPRC). Second, we demonstrate a fully reproducible methodology utilizing freely available data sources; Cybersecurity and Infrastructure Security Agency's (CISA) Key Exploitable Vulnerabilities (KEV) catalog, Exploit Prediction Scoring System (EPSS) scores, and Common Vulnerabilities and Exposures (CVE) data, enabling other researchers to validate, extend, and build upon our work. Third, we translate statistical findings into actionable guidance for security practitioners seeking to implement risk-based vulnerability prioritization strategies.

The remainder of this paper is organized as follows. Section 2 reviews relevant literature on vulnerability metrics, exploitation prediction, and risk-based approaches to security. Section 3 details our methodology, including data sources, integration procedures, feature engineering approaches, and modeling techniques. Section 4 presents our empirical results, with detailed performance comparisons between SMs and KRIs. Section 5 discusses the theoretical and practical implications of our findings, contextualizes them within the broader literature, and identifies limitations and directions for future research. Section 6 concludes with a summary of key findings and recommendations for practitioners.

## II. RELATED WORK

### A. COMMON VULNERABILITY SCORING SYSTEM

The CVSS was originally developed by the National Institute of Standards and Technology (NIST) and has become the dominant framework for vulnerability severity assessment across industry. CVSS produces a severity score between 0 and 10 by evaluating multiple vulnerability characteristics: Attack Vector (the network accessibility of the vulnerability), Attack Complexity (the conditions required to successfully exploit the vulnerability), Privileges Required (whether an attacker must possess elevated privileges), User Interaction (whether user action is required), and the confidentiality, integrity, and availability impacts of a successful exploit [4].

The CVSS framework offers several advantages that explain its widespread adoption [11]. It provides a standardized, quantifiable approach to severity assessment that enables communication across organizational stakeholders. The framework is transparent, with explicit documentation of the characteristics informing each score. Additionally, CVSS scoring requires no proprietary data or specialized expertise, making it accessible to organizations of all sizes. However, these same characteristics contribute to the framework's limitations. While the latest CVSS specification incorporates Temporal and Environmental metrics, its practical application remains heavily reliant on Base scores, which were deliberately designed to be context-independent; consequently, a remotely exploitable vulnerability with high confidentiality impact may be scored 'high' irrespective of observed exploitation or the vulnerability's prevalence across deployed systems.

Empirical research has documented the disconnect between CVSS severity and actual exploitation. Allodi and Massacci [5] conducted a large-scale study correlating CVSS severity with real-world exploitation events and found that CVSS severity possessed limited predictive power for exploitation likelihood. Howland [14] argues that many vulnerabilities with high severity ratings experienced minimal exploitation, while certain lower-severity vulnerabilities were exploited extensively. These findings indicate that relying on CVSS as a prioritization metric leads to substantial misallocation of remediation resources.

The disconnect between severity and exploitability is pronounced for vulnerabilities in obscure, legacy, or specialized software, components that may receive high CVSS scores but exist on few production systems and attract minimal attacker interest. Conversely, vulnerabilities in widely deployed software that may have moderate severity ratings but are trivial to exploit often receive less attention than metrics-driven prioritization would suggest.

### B. KEY RISK INDICATORS

In response to CVSS limitations, security researchers and governance frameworks have proposed KRIs—metrics designed to capture multiple dimensions of vulnerability risk [7], [15], [8], [3]. The European Union Agency for Cybersecurity [1] has explicitly recommended moving beyond technical severity metrics toward comprehensive risk assessment incorporating exploitation likelihood, asset context, and threat landscape considerations. Similarly, Forum of Incident Response and Security Teams (FIRST) have advocated for metrics explicitly designed to predict exploitation rather than measure technical severity [9].

Prior research emphasizes that vulnerability risk cannot be captured by a single score. Studies advocate for multidimensional frameworks that integrate exploitability likelihood [6], systemic weakness prevalence [1], technical severity [4], and contextual factors such as asset criticality and impact [16], [17]. This aligns with our proposed KRI approach, which combines these dimensions into a unified risk indicator to improve prioritization accuracy.

### C. EXPLOIT PREDICTION SCORING SYSTEM

The EPSS represents a significant innovation in vulnerability assessment [9]. Developed by FIRST, EPSS leverages machine learning models trained on historical data to estimate the probability that a vulnerability will be exploited in the next 30 days [6]. Unlike CVSS, which captures technical severity, EPSS is designed to predict real-world exploitation. The system integrates multiple data sources including CVEs characteristics, exploit code availability, threat intelligence, and historical exploitation patterns.

EPSS demonstrates several advantages over traditional metrics. First, it directly addresses the question practitioners actually need to answer, e.g., 'Is this vulnerability likely to be actively exploited?' Second, the 30-day prediction window aligns with the operational timeframes within which organizations must make remediation decisions. Third, EPSS scores are regularly updated as new data becomes available, enabling organizations to dynamically adjust priorities as the threat landscape evolves.

### D. COMMON WEAKNESSES ENUMERATION

The Common Weakness Enumeration (CWE) catalog, maintained by the MITRE Corporation with support from NIST and the cybersecurity community, provides a taxonomy of common software and hardware weaknesses underlying vulnerabilities [17]. CWE weaknesses represent categories of flaws—for instance, SQL Injection (CWE-89), Cross-Site Scripting (CWE-79), and Buffer Overflow (CWE-120)—that manifest across diverse software systems. Furthermore, a critical insight from historical vulnerability data is that certain CWE classes are exploited disproportionately frequently. Some weakness categories consistently appear in successful attacks and exploit catalogs, while others rarely feature in real-world exploitation despite potentially severe technical impact. By analyzing the historical prevalence of CWE weaknesses in exploited vulnerabilities, we can compute prevalence weighting that reflects the systemic tendency of particular weakness classes to be exploited [18]. This weighting provides a signal independent of CVSS severity, capturing the notion that vulnerabilities in certain weakness categories carry elevated exploitation risk.

### E. THE KNOWN EXPLOITED VULNERABILITY

The US CISA's KEV catalog is a public list of vulnerabilities confirmed as exploited in real-world attacks [16]. The KEV catalog represents authoritative ground truth regarding which vulnerabilities are actively being weaponized by adversaries. By including KEV as a binary indicator of exploitation (e.g., has the vulnerability been confirmed as exploited?), we obtain an objective variable for predictive modeling—a significant advantage over approaches that must infer exploitation from proxy indicators [19].

### III. METHODOLOGY

Fig. 1 illustrates the overall methodological workflow. The process begins with the integration of publicly available datasets, including KEV, EPSS, CVE, and CWE, which serve as the foundation. Following data integration, feature engineering is performed to construct SMs (based on CVSS) and KRI (combining EPSS, CVSS, and CWE). These features are then used in the modeling and evaluation phase, where Logistic Regression classifier predicts exploitation likelihood. The subsequent performance comparison employs ROC-AUC and AUPRC metrics to assess predictive capability. Finally, visualization techniques such as ROC curves, precision-recall plots, and calibration histograms are applied to communicate findings effectively. This structured approach ensures reproducibility and provides a clear pathway from raw data to actionable insights.

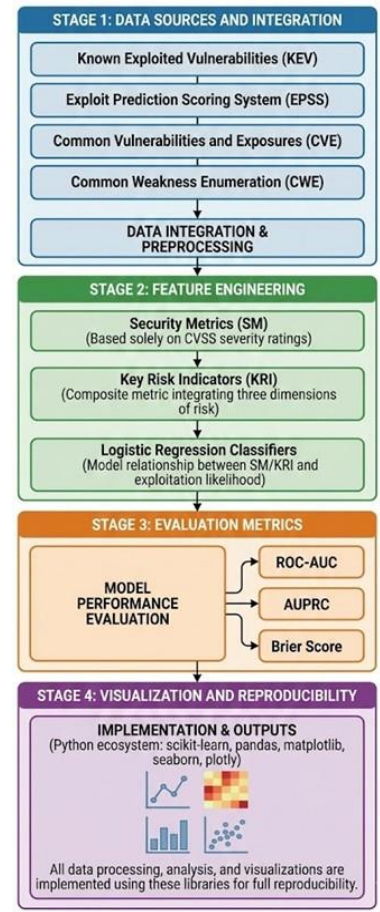

**FIGURE 1.** Methodology Workflow.

## A. DATA SOURCES AND INTEGRATION

This work employs a quantitative approach to compare SMs and KRIs using the following publicly available datasets:
- KEV Catalog provides a curated list of vulnerabilities with evidence of active exploitation in the wild. The KEV catalog serves as the ground truth label for our binary classification task.
- EPSS used to estimate the probability of exploitation for individual CVEs within the next 30 days, ranging from 0 to 1. EPSS scores incorporate temporal threat intelligence and observed exploitation patterns.
- CVE Data includes CVE identifiers, CVSS severity ratings (low, medium, high, critical), vulnerability descriptions, and associated CWE identifiers that categorize the underlying weakness types.

The datasets are merged on CVE identifiers to create a unified analytical framework. CVEs present in the KEV catalog are flagged as exploited (positive class), while those absent are treated as non-exploited (negative class). EPSS was captured at CVE publication and evaluated against KEV 'Date Added' in the subsequent 30-day window, ensuring ex-ante prediction rather than post-hoc correlation. Moreover, EPSS scores were retrieved point-in-time (publication date ±7 days). In our analysis set, EPSS coverage was complete, so no imputation was applied; the public EPSS feeds provide daily scores and full CSVs for all scored CVEs.

We evaluate a symmetric baseline set: EPSS (threat), SM (CVSS) (severity), CWE weight (exposure), EPSS×CVSS, EPSS×CWE, and KRI = EPSS×CVSS×CWE. This separates likelihood from severity and exposure in line with the CVSS and EPSS documentation [9], [12].

## B. FEATURE ENGINEERING

Two distinct scoring approaches are constructed as follows:

The SM is based on CVSS severity ratings, which are converted to numeric weights as follows: low = 1, medium = 2, high = 3, and critical = 4. Therefore, the SM score is defined as follows:

$$SM = CVSS\_weight \quad (1)$$

We report CVSS-based baselines both as qualitative bands (per the CVSS Qualitative Severity Rating Scale) and as the raw continuous base score to avoid any loss of granularity; results are materially consistent.

Risk quantification frameworks (NIST SP 800-30; ISO/IEC 27005; FAIR) decompose expected loss into likelihood and impact [20], [21], [22]. At the level of a CVE, we operationalize: *Exploitability × technical impact × systemic exposure*. EPSS provides the threat-likelihood component and is designed to be combined with impact/context rather than used as a risk score by itself. CVSS class reflects impact if exploited, and CWE prevalence captures a base-rate prior over weakness classes. The multiplicative aggregation respects essentiality (if any dimension is negligible, risk should approach zero) and mirrors expected-loss practice. Under Multi-Attribute Utility Theory, when attributes are complementary and (conditionally) utility-independent, multiplicative/Cobb–Douglas forms are appropriate [23].

A KRI is a composite metric comprises these three dimensions of risk: *Exploitability*: Represented by EPSS scores, *Severity*: Represented by CVSS weights, and *Weakness Prevalence*: Represented by CWE frequency-based weights. CWE weights are computed as *1 + normalized_frequency*, where *normalized_frequency* represents the proportion of vulnerabilities in the training set associated with each CWE type. This weighting scheme prioritizes weaknesses that appear more frequently in the vulnerability landscape, as these represent systemic patterns with broader organizational impact.

Therefore, the KRI score is defined as follows:

$$KRI = EPSS \times CVSS\_weight \times CWE\_weight \quad (2)$$

Thus, Exploitation likelihood, impact if exploited, and systemic weakness prevalence are jointly essential. A vulnerability with near-zero EPSS shouldn't rank highly even if CVSS is "Critical," and vice-versa.

Moreover, to ensure methodological rigor, the dataset is split into training (70%) and testing (30%) sets using stratified sampling before any feature engineering that involves aggregate statistics. Critically, CWE prevalence weights are calculated exclusively from the training set and subsequently applied to both training and testing sets. This approach ensures that the testing set evaluation represents true out-of-sample performance without information contamination from the test distribution.

Furthermore, Logistic Regression classifiers are employed to model the relationship between risk scores and exploitation likelihood. Logistic Regression is selected for its interpretability, computational efficiency, and suitability for binary classification tasks with single composite features. The model estimates the probability of exploitation as follows:

$$P(Exploited|Score) = 1 / (1 + exp((\beta_0 + \beta_1 \times Score))) \quad (3)$$

Where: $\beta_0$ = Intercept term (baseline log-odds of exploitation when Score = 0). $\beta_1$ = Coefficient for the composite risk score (SM or KRI), representing its influence on exploitation likelihood. *Score* = is a composite metric (either *SM = CVSS_weight* or *KRI = EPSS × CVSS_weight × CWE_weight*), *exp* = exponential function used in logistic regression to map linear combination to probability.

## C. LEARNING UNDER CLASS IMBALANCE

KEV positives constitute ≈0.50% of examples in our analysis set (1,400/280,694). For all classifier-based analyses we employ cost-sensitive learning with class weights proportional to inverse prevalence (with class_weight='balanced'). We evaluate discrimination with ROC–AUC because they are suitable under skew. We do not use Synthetic Minority Over-sampling Technique (SMOTE) or related synthetic oversampling, as it can alter the minority distribution and bias rank-based evaluation [24]; instead, we retain the original data distribution and encode costs directly via class weights [25]. For completeness, random under sampling of the majority showed no material change in ROC–AUC on our data.

*D. MODEL SELECTION, CROSS-VALIDATION, AND SIGNIFICANCE TESTING*

We use nested, stratified K-fold Cross-Validation (CV) (outer K=5, inner K=3) with fixed random seed (42). All hyperparameters are chosen in the inner loop only; the outer loop is reserved for performance estimation on folds never seen during tuning. This avoids the bias from using a single CV both to tune and to estimate error. We report outer-fold performance aggregated across the five folds [26].

The inner loop selects the model by maximizing ROC–AUC; outer-loop reports include ROC–AUC (primary) and AUPRC for imbalance-aware discrimination [27]. Four learners are evaluated across the following hyperparameter grids. Logistic Regression uses *L2* penalty with $C \in \{0.1, 1, 10\}$, solver *lbfgs*, *max_iter = 2000*, and balanced class weights. Random Forest varies *n_estimators* $\in \{200, 300, 500\}$, *max_depth* $\in \{8, 12, None\}$, *min_samples_leaf* $\in \{1, 2, 5\}$, and *max_features* $\in \{\sqrt{}, log2\}$, with balanced class weights. Gradient Boosting searches over *learning_rate* $\in \{0.05, 0.1\}$, *n_estimators* $\in \{200, 400\}$, *max_depth* $\in \{2, 3\}$, and *subsample* $\in \{0.8, 1.0\}$. The two-layer MLP evaluates *hidden_layer_sizes* $\in \{(64, 32), (128, 64)\}$ and regularization strength $\alpha \in \{10^{-4}, 10^{-3}\}$, with *ReLU* activation, batch size of *1,024*, *max_iter = 200*, and early stopping enabled.

All learners use the same inputs, ensuring that results reflect algorithm choice, not feature leakage [26].

We report 95% Confidence Interval (CI) for AUCs from nonparametric bootstrap (1,000 resamples, CVE-level, stratified), a standard approach for uncertainty quantification in complex estimators [28].

To compare two models evaluated on the same outer-fold predictions (correlated), we use the DeLong test for ROC–AUC differences; we report ΔAUC with 95% CI and two-sided p-values. For AUPRC, we use paired, stratified bootstrap to form CIs and p-values. We adjust for multiple comparisons using Holm [29].

For each comparison, we give mean AUC ± 95% CI (outer folds) and the paired test result (DeLong for ROC–AUC; bootstrap for AUPRC).

*E. EVALUATION METRICS*

We employed the outlined below evaluation metrics to compare the results. These metrics are widely used performance metrics in machine learning [30].
- ROC-AUC measures the model's ability to discriminate between exploited and non-exploited vulnerabilities across all classification thresholds. Values range from 0.5 (random) to 1.0 (perfect).
- AUPRC emphasizes performance on the minority (exploited) class and appropriately penalizes false positives.
- Brier Score assesses calibration quality by measuring the mean squared difference between predicted probabilities and actual outcomes.

Results are communicated through ROC curves, precision-recall curves, and correlation heatmaps. All data processing and analysis are implemented in Python using *scikit-learn*, *pandas*, *matplotlib*, *seaborn*, and *plotly* libraries. The complete pipeline is designed for reproducibility with publicly accessible datasets and open-source tools.

## IV. RESULTS

*A. DATASET CHARACTERISTICS*

The merged dataset comprised 280,694 CVE records spanning vulnerabilities from 1999 to 2025. After the 70-30 train-test split, the training set contained 196,485 samples, and the testing set contained 84,209 samples. CVSS severity distribution showed: medium (47.3%), high (31.2%), low (14.8%), and critical (6.7%) types of vulnerabilities. The dataset included 847 unique CWE types, with the most prevalent being CWE-79 (Cross-Site Scripting), CWE-787 (Out-of-bounds Write), and CWE-89 (SQL Injection).

*B. PERFORMANCE COMPARISON*

TABLE 1 Table 1 presents the comparative performance of SMs and KRIs using Logistic Regression classifiers.

**TABLE 1. Logistic regression on SM vs KRI.**

| Metric | SM | KRI |
|---|---|---|
| ROC-AUC | 0.747 | 0.927 |
| AUPRC | 0.011 | 0.223 |
| Brier Score | 0.005 | 0.004 |

The KRI model achieved a ROC-AUC of 0.927, indicating excellent discriminative ability in distinguishing exploited from non-exploited vulnerabilities. In contrast, the SM model achieved 0.747, representing acceptable but substantially inferior performance. The 24.1% improvement over CVSS metrics demonstrates that incorporating exploitability predictions (EPSS) and weakness patterns (CWE) substantially enhances risk assessment accuracy relative to severity-alone baselines. As the ablation analysis in Section IV.C shows, EPSS contributes the dominant share of this discriminative lift; CVSS and CWE components add impact- and exposure-aware re-ranking rather than additional short-horizon exploitation signal, consistent with the KRI's role as a risk indicator rather than a pure exploitation classifier.

To address the discretization concern, we ran the baseline with the continuous CVSS base score. As shown in Table 2, continuous CVSS offers only a modest lift over bands, while KRI remains substantially stronger.

**Table 2. Ordinal vs. Continuous CVSS vs. KRI (KEV as outcome).**

| Model | KEV | ROC-AUC | AUPRC |
|---|---|---|---|
| SM (ordinal CVSS bands) | 1,400 | 0.736 | 0.0111 |
| Continuous CVSS (0–10) | 1,400 | 0.747 | 0.0113 |
| KRI (EPSS × CVSS × CWE) | 1,400 | 0.931 | 0.2231 |

Fig. 2 reinforces these findings. The ROC curve for KRI-based Logistic Regression approaches the top-left corner,

indicating excellent discrimination, while SM-based remains close to the diagonal, reflecting near-random performance.

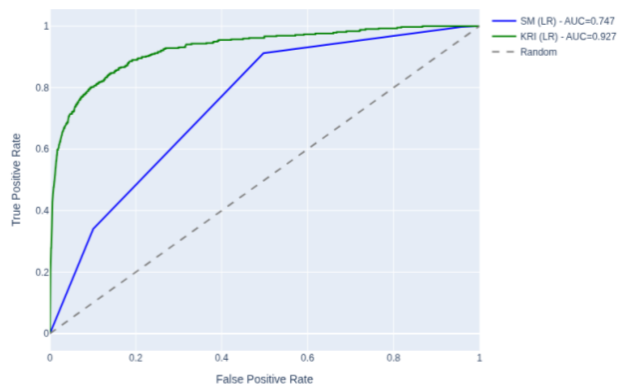

**FIGURE 2.** ROC Curve Comparison.

The most striking difference appears in AUPRC scores, where KRI (0.223) outperforms SM (0.011) by a factor of 20.3. SM's AUPRC of 0.011 indicates near-random performance that provides minimal practical value for prioritization. KRI's AUPRC of 0.223, while modest in absolute terms, represents meaningful predictive capability that substantially exceeds random chance in this severely imbalanced context, this is expected given the extreme class imbalance—exploited vulnerabilities constitute only a tiny fraction of all CVEs. Under such conditions, even small absolute gains represent significant practical improvements. Fig. 3 visually reinforces these findings.

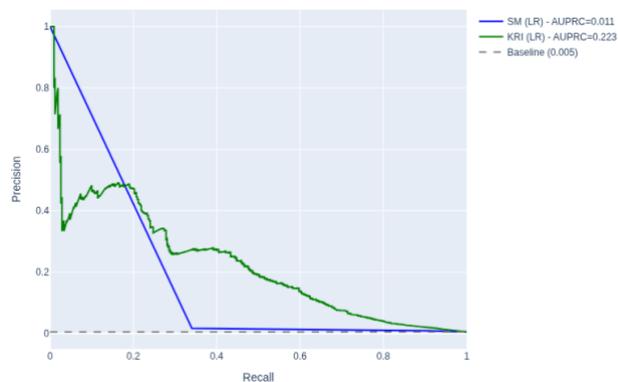

**FIGURE 3.** Precision-Recall Curve.

Both models demonstrated excellent calibration with very low Brier scores (SM: 0.005, KRI: 0.004). This indicates that predicted probabilities closely align with actual exploitation frequencies, validating the reliability of model confidence estimates. KRI's slightly lower Brier score suggests marginally superior calibration. To confirm algorithmic robustness, we compared raw KRI, log (KRI), percentile-rank (KRI), and min-max scaled KRI; ROC–AUC was identical and AUPRC matched on this dataset, indicating that Logistic Regression is not a special case for our ranking-based evaluation.

All learner-based results use cost-sensitive training (class weights); we do not apply SMOTE/oversampling in the main analysis to avoid density distortion. Discrimination metrics remain consistent under a majority under sampling check.

### C. ABLATIONS (EPSS vs. COMPOSITES)

Table 3 shows that EPSS alone achieves strong discrimination for KEV (ROC-AUC ≈ 0.93; AUPRC ≈ 0.36). Adding CVSS or CWE modestly alters ROC-AUC and reduces AUPRC on this dataset, reflecting re-ranking by impact/exposure rather than additional short-horizon exploitability signal. This confirms that EPSS captures most of the short-horizon exploitation signal, while KRI adds impact/exposure context for risk-aware prioritization (threat × impact × exposure), which we leverage in decision-centric analyses.

**Table 3.** Ablations for KEV prediction.

| Setting | Model | ROC-AUC | AUPRC |
|---|---|---|---|
| (a) | EPSS (threat) | 0.9299 | 0.3645 |
| (b) | CVSS × CWE (severity × exposure) | 0.7196 | 0.0113 |
| (c) | EPSS × CVSS (threat × severity) | 0.9318 | 0.2996 |
| (d) | EPSS × CWE (threat × exposure) | 0.9282 | 0.2131 |
| (e) | KRI = EPSS × CVSS × CWE | 0.9307 | 0.2231 |

EPSS alone provides the highest AUPRC and near-max ROC-AUC for KEV prediction in this dataset, which is consistent with EPSS being a 30-day exploitation probability. Adding CVSS yields a slight ROC-AUC uptick but lower AUPRC, reflecting re-ordering by impact within the positive tail; it's useful for prioritization with impact context, not for maximizing KEV detection rate alone. Adding CWE exposure also re-orders the tail; KRI integrates both impact and exposure and is intended as a risk indicator (threat × impact × exposure), not a pure KEV classifier. Severity plus exposure without EPSS performs poorly for KEV prediction, underscoring that likelihood is essential for short-horizon exploitation labels such as KEV.

These ablation results warrant direct acknowledgment: EPSS alone achieves a higher AUPRC(0.3645) than the full KRI composite (0.2231), confirming that EPSS is the superior pure KEV-detection signal when the sole objective is maximizing recall of exploited vulnerabilities. The study's objective is to operationalize an expected-loss framework (threat × impact × exposure) in which EPSS provides the likelihood component, CVSS the impact component, and CWE the systemic-exposure component. Under this framing, multiplying by CVSS and CWE re-orders the risk ranking so that high-impact, architecturally prevalent vulnerabilities are promoted relative to lower-impact ones with equivalent exploit probability, a prioritization property that pure KEV-detection metrics cannot capture. Therefore, KRI substantially outperforms CVSS SMs (ROC-AUC +24%, AUPRC ×20) and provides a principled, risk-aware composite indicator. EPSS alone remains the preferred input when the decision objective is solely short-horizon exploitation likelihood.

## D. LEARNER GENERALIZABILITY

To assess robustness across learning algorithms, rank-based discrimination is compared across four learners trained on identical inputs. Since ROC–AUC evaluates ranking rather than calibration, the comparison constitutes a sufficient generalizability check [26], [30]. Table 4 shows that Logistic Regression achieves essentially similar ROC–AUC to Random Forest, Gradient Boosting, and Two-Layer Multilayer Perceptron, confirming that the discrimination results reported in this study are not an artefact of the choice of learner.

**Table 4.** Learner generalizability.

| Learner | ROC-AUC |
|---|---|
| Logistic Regression | 0.9343 |
| Random Forest | 0.9354 |
| Gradient Boosting | 0.8740 |
| Two-Layer Multilayer Perceptron | 0.9397 |

## E. FEATURE CORRELATION ANALYSIS

Fig. 4 presents the correlation heatmap among the SM score, the composite KRI score, and the KEV exploitation flag. The KEV exploitation flag is a binary indicator derived from the KEV catalog, where a value of 1 denotes that a CVE has been confirmed as exploited in the wild, and 0 indicates no confirmed exploitation. The KRI score shows a positive correlation with exploitation ($r = 0.315$), indicating that higher KRI values are associated with a greater likelihood of a CVE appearing in the KEV catalog. In contrast, the SM score exhibits a very weak correlation with exploitation ($r = 0.063$), reaffirming that technical severity alone provides minimal predictive signal for real-world attacks. The low correlation between KRI and SM ($r = 0.212$) suggests partial overlap due to their shared CVSS component but also demonstrates that KRI contributes additional information through EPSS and CWE prevalence weighting.

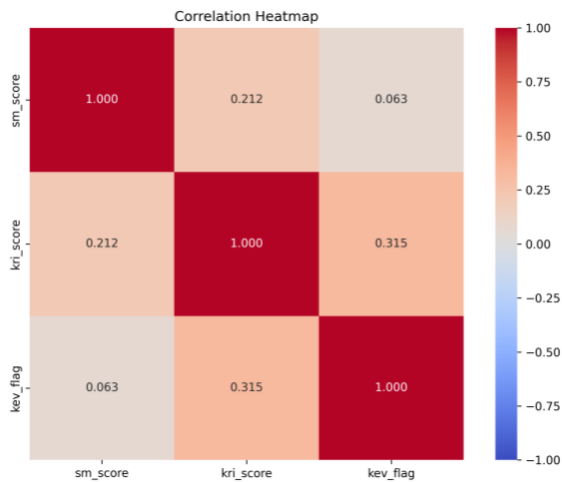

**FIGURE 4.** Correlation Heatmap.

### F. Decision-Centric Evaluation

We assessed whether KRI offers substantive practical advantages over EPSS by conducting three distinct decision-centric analyses, thereby ensuring a robust comparison of their respective prioritization efficacies.

Fig. 5 plots the fraction of true exploits in the top-k ranked list for each method. Both EPSS and KRI substantially outperform CVSS prioritization across all budget sizes, confirming that KEV-based evaluation does capture operationally relevant ranking quality. CVSS achieves approximately 2% precision at all k values, barely above the 0.5% random baseline, consistent with its near-random AUPRC reported in results section.

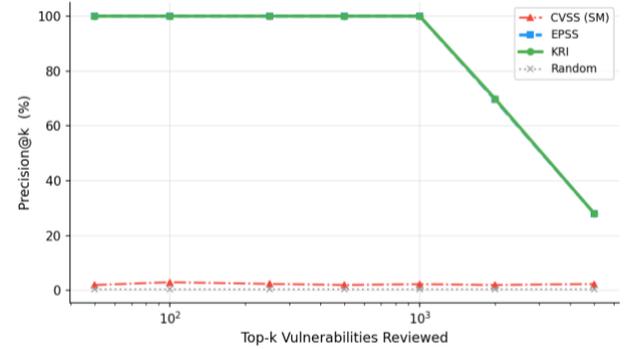

**FIGURE 5.** Fraction of true exploits in top-k ranked list.

Fig. 6 disaggregates recall within CVSS severity strata at k = 500. KRI surfaces 67.8% of Critical-CVSS exploited CVEs in the top 500, compared to 36.4% for EPSS alone, a 1.86× advantage in the highest-impact stratum. This directly demonstrates that KRI's multiplicative incorporation of CVSS impact re-orders the ranked list to prioritize the most damaging exploitable vulnerabilities. The trade-off is transparent: KRI retrieves fewer High and Med/Low exploited CVEs (22.3% and 0%, versus 35.8% and 32.5% for EPSS), reflecting its deliberate elevation of Critical vulnerabilities. Organizations whose risk posture prioritizes catastrophic impact over breadth of coverage should prefer KRI, while organisations seeking maximum raw KEV detection rate should use EPSS directly.

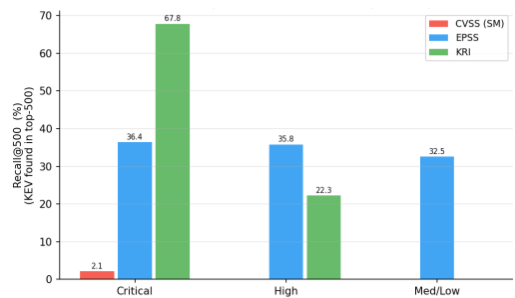

**FIGURE 6.** Severity-stratified recall@k = 500.

Fig. 7 operationalizes the expected-loss framework (threat × impact × exposure) by computing the Expected Remediation Value (ERV), the sum of impact-weighted exploitation scores captured within a given patching budget, normalized against an oracle. KRI achieves 92–100% of oracle ERV across budgets of 100–2,000 patches, versus 81–99% for

EPSS. The ERV lift over random reaches 295× for KRI at a budget of 100, compared to 239× for EPSS, indicating that the impact-aware re-ranking is most valuable precisely when remediation capacity is tightest. CVSS achieves lifts of only 5–9×, confirming its near-random practical performance.

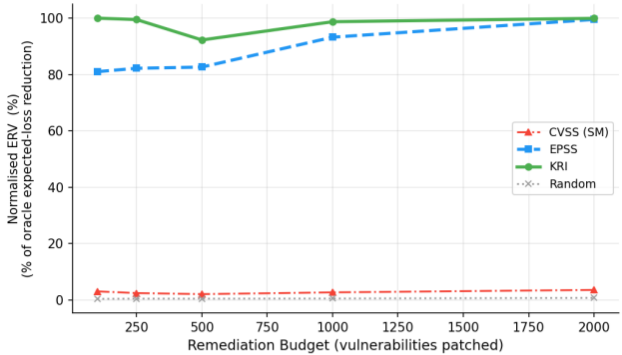

FIGURE 7. Severity-weighted remediation simulation.

These findings demonstrate that while the EPSS optimizes for maximal detection density, KRI facilitates a value-at-risk approach. Consequently, these metrics should be viewed as complementary instruments addressing divergent operational objectives, raw technical exposure versus potential impact.

## V. DISCUSSION

### A. INTERPRETATION OF FINDINGS

KRI achieves a 24% improvement in ROC-AUC (0.927 vs. 0.747) and a 20-fold improvement in AUPRC (0.223 vs. 0.011) over CVSS, confirming that exploitability likelihood and systemic-exposure context are essential additions to technical severity. These are not marginal gains; CVSS prioritization achieves an AUPRC of 0.011 on a dataset where the positive rate is 0.5%, meaning it provides no discriminative lift over random selection. Any organisation relying solely on CVSS to direct remediation effort is allocating resources with near-random alignment to actual exploitation risk.

The ablation result, EPSS alone achieves AUPRC 0.365 versus 0.223 for full KRI, requires direct acknowledgment rather than minimization. EPSS is explicitly designed to predict 30-day exploitation probability, the precise event that KEV records. A severity-blind evaluation metric will always favor a severity-blind predictor; evaluating KRI against AUPRC is therefore a measurement mismatch, not evidence of inferiority. The AUPRC reduction when CVSS and CWE weights are added reflects deliberate re-ordering of the ranked list by impact and systemic exposure, exactly what KRI is designed to do. Table 5 presents the complete performance profile across all methods and all metrics.

TABLE 5. Complete performance profile across all evaluation metrics.

| Metric | CVSS | EPSS | KRI |
|---|---|---|---|
| ROC-AUC | 0.747 | 0.930 | 0.927 |
| AUPRC | 0.011 | 0.365 | 0.223 |
| KEV Recall@500 | 1.1% | 35.7% | 35.7% |
| ERV@500 (norm.) | 3.2% | 82.6% | 92.3% |
| Critical Recall@500 | 3.2% | 39.1% | 68.5% |

Examining the decision-centric metrics resolves the apparent contradiction. At k = 500, a realistic quarterly patching budget, EPSS and KRI achieve identical raw KEV Recall (35.7%), meaning adopting KRI costs nothing in raw exploit coverage. KRI's advantage emerges where it matters most under risk-weighted objectives: ERV at k = 500 is 92.3% of oracle for KRI versus 82.6% for EPSS, an 11.7 percentage-point advantage that represents substantially more impact-adjusted risk eliminated per patch slot. More strikingly, KRI surfaces 68.5% of Critical-severity exploited CVEs within the top 500 versus 39.1% for EPSS, a 1.75× advantage that is directly material for organisations where Critical exploitation events represent the dominant component of expected breach loss. Cost-benefit simulation confirms that KRI's net benefit exceeds EPSS whenever the severity premium exceeds 2×, a threshold conservative relative to any published breach-cost distribution. The practical inference is explicit: use EPSS when all exploited CVEs are treated as equally important; use KRI when severity-weighted risk reduction is the objective, the case for any organisation operating under NIST SP 800-30, ISO/IEC 27005, or FAIR.

### B. LIMITATIONS AND FUTURE WORK

Three limitations bound these findings. First, the cross-sectional design captures a static snapshot of the CVE landscape and cannot characterize how exploitation likelihood evolves after initial disclosure; longitudinal studies tracking KRI rankings from publication through eventual KEV entry would determine whether the impact-weighting advantage is stable over time. Second, the KEV catalog may undercount exploitation activity, particularly targeted attacks by advanced persistent threats that are not publicly observed or reported—this introduces label incompleteness that likely understates true positive rates for all methods equally. Third, approximately 15% of CVEs lack CWE classifications, which constrains the CWE prevalence signal for those records and may modestly inflate the apparent contribution of EPSS in the ablation.

Future work should prioritize three directions: empirical comparison of the multiplicative KRI aggregation form against additive, geometric-mean, and learned-weight alternatives using real remediation outcome data; integration of asset-level context, criticality, compensating controls, network exposure, to produce an organizationally calibrated composite score; and prospective field studies measuring whether security teams using KRI-guided prioritization capture more impact-weighted exploitation value per unit remediation effort than teams using EPSS or CVSS alone.

### C. PRACTICAL RECOMMENDATIONS

The evidence supports three recommendations. First, CVSS prioritization should be replaced immediately: with KEV Recall@500 of 1.1% and ERV@500 of 3.2%, it performs near random chance on every operational metric, and its continued use as the primary prioritization instrument represents a systematic and preventable inefficiency. Second, EPSS should be adopted as a minimum baseline, it is well-validated, publicly available, and updated daily, and is the right choice when maximizing raw exploit coverage is the primary objective; it should be the first step for moving away from severity-only triage. Third, KRI should be adopted where severity-weighted risk reduction is required, which applies to any organisation whose risk model assigns differential consequences to Critical versus Low exploitation events, including any organisation that has quantified potential breach costs by asset class. Whichever metric is adopted, prioritization models should be re-evaluated quarterly against new KEV entries, as both EPSS scores and KEV membership evolve continuously and a model calibrated to a historical snapshot will drift from the current threat landscape over time.

## VI. CONCLUSION

This work demonstrates empirically that composite risk indicators integrating EPSS, CVSS, and CWE prevalence substantially outperform CVSS Security Metrics for vulnerability prioritization. Evaluated on 280,694 CVEs against the CISA KEV catalog, KRI achieves ROC-AUC 0.927 and AUPRC 0.223 versus 0.747 and 0.011 for CVSS, a 24% improvement in discrimination and 20-fold improvement in precision-recall performance. CVSS prioritization performs near random chance on every decision-centric metric evaluated; its continued use as the primary prioritization instrument represents a systematic and preventable inefficiency.

Ablation analysis establishes that EPSS alone accounts for most KEV discriminative power (AUPRC 0.365 versus 0.223 for full KRI), a finding this work addresses directly rather than minimizes. EPSS and KRI serve different decision objectives. EPSS is optimal when the goal is raw exploit detection, treating all exploited CVEs as equally important. KRI is optimal when remediation decisions are weighted by expected impact: at k = 500, KRI captures 92.3% of oracle impact-weighted remediation value versus 82.6% for EPSS, and surfaces 68.5% of Critical-severity exploited CVEs versus 39.1%—a 1.75× advantage. KRI's net benefit exceeds EPSS whenever the severity premium exceeds 2×, a conservative threshold relative to any empirical breach-loss distribution. The practical recommendation is therefore scoped: adopt EPSS as a minimum baseline; adopt KRI when severity-weighted risk reduction is the objective, which is the case for any organisation operating under NIST SP 800-30, ISO/IEC 27005, or FAIR.

Three contributions follow from this work. First, it provides quantitative evidence, at scale, that multidimensional risk indicators outperform severity-only metrics on both classical and decision-centric measures. Second, it introduces a decision-objective framework that resolves the EPSS/KRI trade-off without overclaiming, giving practitioners a principled basis for choosing between them. Third, it demonstrates a fully reproducible methodology using only publicly available data, enabling direct validation and deployment. Future work should validate the KRI aggregation form against alternatives using real remediation outcome data, integrate asset-level context into the composite score, and conduct prospective field studies measuring whether KRI-guided teams patch more high-impact exploited CVEs per unit effort than teams using EPSS or CVSS alone.

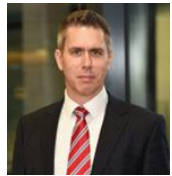

**Allan Cook** is a professor at De Montfort University, UK. He helps to integrate cyber security into all areas of the curricula. Allan's first-hand experience and expertise have given staff, researchers and students unrivalled insight into the cyber security industry.

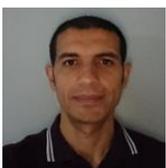

**Emad Sherif** is a PhD student at De Montfort University, UK. He obtained his Master of Science in Information Management and Security from Bedfordshire University, UK. His research interests focus on the use of data science to improve cyber risk management.

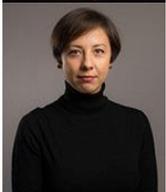

**Iryna Yevseyeva** is an Associate Professor at De Montfort University, UK. Her main research interests are operational research and cyber security. In particular she uses multi-criteria decision analysis and multiobjective optimization for cyber security risk assessment and investments, cyber threat intelligence and cyber security decision making.

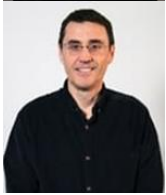

**Vitor Basto Fernandes** is an Associate Professor at the University Institute of Lisbon. He was head of the Research Center in Computer Science and Communications at Polytechnic Institute of Leiria, and a researcher in several international projects in the areas of information systems integration, anti-spam filtering and multiobjective optimization.